\documentstyle[11pt,paspconf,epsf]{article}

\begin{document}

\title{The Results From The NICMOS Parallel Imaging and Grism Survey}

\author{Lin Yan}
\affil{The Carnegie Observatroies, Pasadena, CA91108, email: lyan@ociw.edu}

%
%




\begin{abstract}
We present the results of a survey which utilizes the NICMOS
Camera 3 Parallel grism and imaging observations of random
fields. We have identified $33$ H$\alpha$ emission-line galaxies at $ 0.75 < z < 1.9$.
The inferred co-moving number density of these objects
is $3.3\times10^{-4}~h_{50}^{3}$~Mpc$^{-3}$, very
similar to that of the bright Lyman break objects at $z \sim 3$. 
The mean star formation rate of these galaxies 21 M$_{\odot}$ yr$^{-1}$ for
H$_0=50$km/s/Mpc.
Using this sample, we derived the H$\alpha$ luminosity function (LF) at $z=1.3$.
The integrated H$\alpha$ luminosity density at $z \sim 1.3$ is
$1.64\times 10^{40}$~h$_{50}$~erg s$^{-1}$ Mpc$^{-3}$, approximately 14 times
greater than the local value reported by Gallego et al. (1995).
The volume averaged star formation rate
at $z = 1.3 \pm 0.5$ is 0.13 M$_{\odot}$ yr$^{-1}$ Mpc$^{-3}$ without correction
for extinction. The SFR derived at $\sim 6500$~\AA\
is a factor of $3$ higher than that deduced from 2800~\AA\ continua.
We believe that this difference is largely due to dust extinction. The implied
total extinction at 2800\AA\ is in the range of $2 - 4$ magnitude. 
However, the precise determination of the total extinction is sensitive to the model assumptions. 

Deep ground-based VRI images of the NICMOS fields have revealed
roughly a dozen of extremely red objects (EROs) with $\rm R - H >5$ and $\rm H$
brighter than 20.6. The surface density of these objects is around 0.6 per square arcminutes.

\end{abstract}


\keywords{Emission Line Objects; High redshift galaxies; Star formation rates}


\section{Introduction}

NICMOS parallel observations, taken in parallel with one of the other
science instruments on HST, has provided us for the first time
a wealth of data at near  infrared wavelengths with HST resolution.
Small background at wavelengths of 0.8$\mu$ and 1.6$\mu$ and HST
high angular resolution make NICMOS a very efficient instrument
in studying the faint galaxy population at high redshifts.

The NICMOS parallel imaging and grism observations were both made
with Camera 3 with a field of view $\sim$ 52$''\times$ 52$''$.
The imaging data were taken with broad band filters F110W and F160W
at 1.1$\mu$ (J band) and 1.6$\mu$ (H band).
The grism data has a spectral resolution of 200 per pixel and
covers wavelength regions from 0.8$\mu$ to 1.2$\mu$ (G096)
and 1.1$\mu$ to 1.9$\mu$ (G141). We have reduced and analysed 
the NIC3 parallel imaging data covering $\sim$150 sq. arcminutes
and the grism data in G141 grism $\sim$65 sq. arcminutes.

\section{Emission-line Galaxies and the H$\alpha$ Luminosity Function at
$z \sim 1.3$ from the NICMOS/HST Grism Parallel Observations}

The recent detections
of dust enshrouded galaxies at $z > 1$ at sub-millimeter
wavelengths (Smail, Ivison \&\ Blain 1997; Hughes et al. 1998;
Barger et al. 1998; Lilly et al. 1998) suggest that significant amounts of star
formation activity at high redshifts may be obscured. Observations in 
the rest-frame UV wavelength suffer large uncertainties in
extinction corrections. Furthermore, 
little is known about the properties of normal galaxies in
the region between $ 1 < z < 2$, where neither the 4000\AA\ break nor
the Ly continuum break are easily accessible.  
The near-IR offers one means of accessing both redshift indicators and
measures of star formation within this critical redshift range.

We have reduced and analysed the NIC3 parallel grism G141 data, covering
$\sim$ 65 sq. arcminutes.
The details of the data reduction can be found in McCarthy et al. (1999).
We found a total of 33 emission line galaxies over an 
effective co-moving volume of $10^5~h_{50}^{-3}$~Mpc$^3$ for $q_0=0.5$.
The implied co-moving number density of emission line galaxies in the range
$0.75 < z <  1.9$ is $3.3\times10^{-4}~h_{50}^{3}$~Mpc$^{-3}$, very
similar to that of the bright Lyman break objects at $z \sim 3$.  
These objects have a median H$\alpha$ luminosity of $2.7 \times
10^{42}$ erg sec$^{-1}$. The most, if not all, of the 
emission lines detected are either H$\alpha$ or unresolved
blend of H$\alpha$+[NII]6583/6548. This identification is mostly based
on H-band apparent magnitudes, the emission line equivalent widths 
and the lack of other detected lines within the G141 bandpass.
The median H-band apparent magnitude of $\sim$20.5
(which corresponds to a L$^\star$ galaxy at $z \sim 1.5$) implies that 
the possibility of the line being [OII] or H$\beta$ is very small.
The redshifts of 6 galaxies in our sample have been confirmed by detection of 
[OII]3727 emission in the optical spectra using LRIS on the Keck 
10m telescope (Teplitz et al. 1999; Malkan et al. 1999).
The fraction of AGN contamination in our sample is around
10\%; for details, see McCarthy et al. (1999). Figure 1 shows 
the spectra for a subset of galaxies in our sample.

\begin{figure}
\plotfiddle{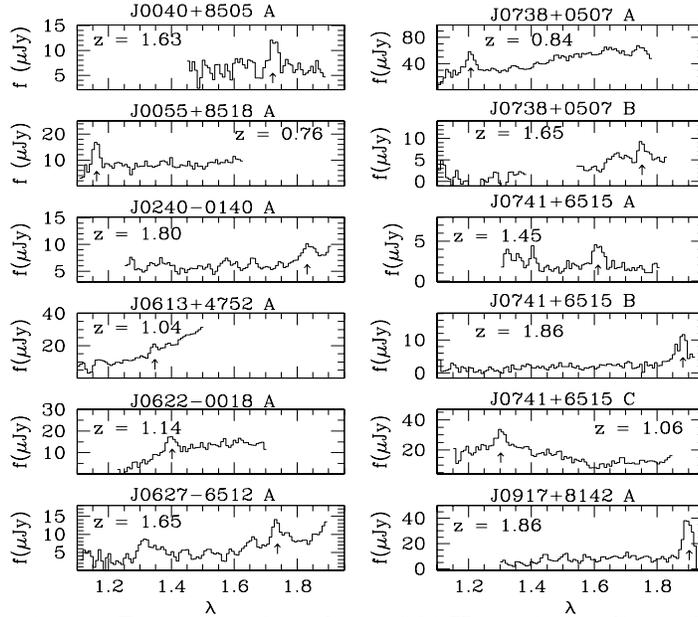}{6.5truecm}{0.0}{50}{45}{-160}{-100}
\caption{1-D spectra of a subset of 33 H$\alpha$ emission-line galaxies.
For each object we have marked our candidate emission-lines
with arrows below the line. We plot the entire range of the G141 grism
for each spectrum even though parts of the spectrum have fallen beyond
the field of view of the detector for several objects. The redshifts,
assuming an H$\alpha$ identification for the line are given for each
object.}
\end{figure}

We compute the H$\alpha$
luminosity function (LF) based on our sample of emission-line galaxies. 
We have corrected the incompleteness in the original source catalog using simulations. 
The final correction is significant only in the faintest luminosity bin and our main result
does not sensitively depend on that. All of the detailed results are in Yan et al. (1999).
Figure 2 shows our derived H$\alpha$ LF at $z=1.3$ and the local H$\alpha$
LF as measured by Gallego et al. (1995).
This plot shows strong evolution in the H$\alpha$ luminosity density
from $z\sim 0 $ to $z \sim 1.3$.  This is no surprise given the evolution in
the ultraviolet luminosity density, but our result provides an independent
measure of evolution for H$\alpha$ emission alone. The LF is well fit
by a Schechter
function over the range $6 \times 10^{41} <$ L$(H\alpha$) $ < 2 \times 10^{43}$
erg sec$^{-1}$ with L$^{*} = 7 \times 10^{42}$ erg sec$^{-1}$
and $\phi^* = 1.7 \times 10^{-3}$ Mpc$^{-3}$ for H$_0=50$~km s$^{-1}$ Mpc$^{-1}$ 
and q$_0=0.5$. The integrated H$\alpha$
luminosity density at $z \sim 1.3$ (our median $z$) is
$1.64\times 10^{40}$~h$_{50}$~erg s$^{-1}$ Mpc$^{-3}$, $\sim$14 times
greater than the local value reported by Gallego et al. (1995).

\begin{figure}
\plotfiddle{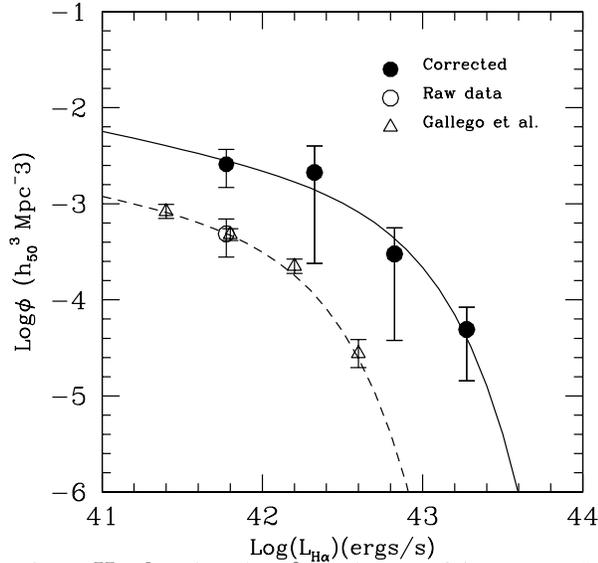}{6.5truecm}{0.0}{45}{45}{-160}{-100}
\caption{H$\alpha$ luminosity function at $ 0.7 < z < 1.9$.
The open and filled circles are the data points from our measurements.
The open circles represent the raw data and the filled circles
are the points corrected for incompleteness. The incompleteness correction
is only significant at the faintest luminosity bin. The open triangles
show the local H$\alpha$ luminosity function by Gallego et al. (1995).
The solid and dashed lines are the best fits to the data
at $z \sim 1.3$ and $z \sim 0$ respectively.}
\end{figure}

We converted the integrated H$\alpha$ luminosity density to a star
formation rate (SFR) using the relation from Kennicutt
(1999):  $\rm {SFR}(M_\odot yr^{-1}) = 7.9 \times 10^{-42} L(H\alpha)
(erg~s^{-1}) $.  This assumes Case B recombination at $T_e =
10^4$~K and a Salpeter IMF ($0.1 - 100~M_\odot$). This conversion
factor is about 10\% smaller than the value listed in Kennicutt (1983),
the difference reflecting updated evolutionary tracks.
While different choices of stellar tracks introduce modest uncertainties
in the conversion of UV and H$\alpha$ luminosities to star formation rates,
the choice of different IMFs lead to rather large differences.
To make consistent comparisons between our results and those in the literature
derived from 1500\AA\ and 2800\AA\ UV continuum luminosity
densities, we adopt the relation from Kennicutt (1999):
$\rm {SFR}(M_\odot yr^{-1}) = 1.4 \times 10^{-28} L(1500-2800\AA)
(erg~s^{-1} Hz^{-1}) $. This relation is appropriate for
the Salpeter IMF used to derive the H$\alpha$ conversion factor.

\begin{figure}
\plotfiddle{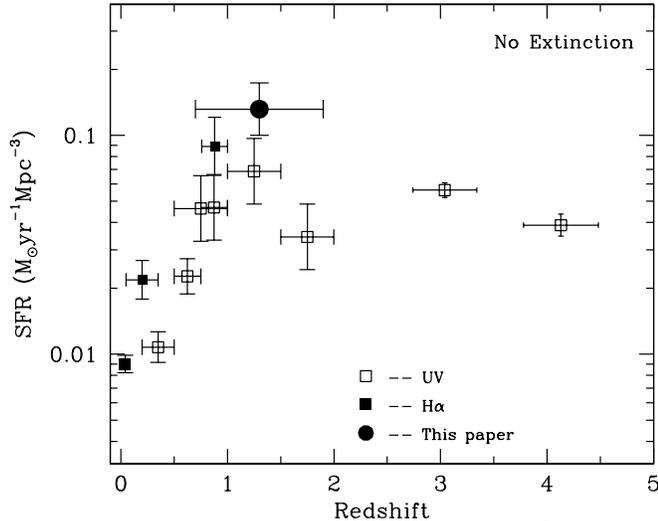}{6.0truecm}{0.0}{45}{45}{-160}{-120}
\caption{The global volume-averaged star formation rate as a function of
redshift without any dust extinction correction. The open squares
represent measurements of the 2800\AA\ or 1500\AA\ continuum luminosity density by
Lilly et al. (1996), Connolly et al. (1997) and Steidel et al.  (1998),
whereas the filled squares are
the measurements using H$\alpha$~6563\AA\ by
Gallego et al. (1995), Tresse \&\ Maddox (1996) and Glazebrook et al. (1998).
Our result is shown in the filled circle.}
\end{figure}

In Figure 3, we plot uncorrected published measurements of the
volume-averaged global star formation rate at various epochs.
Our result is shown as a filled circle.
The star formation rates shown in Figure 3 are calculated from
the luminosity densities integrated over the entire luminosity
functions, for both H$\alpha$ and the UV continuum.
Lilly et al. and Connolly et al. assumed a faint end slope of
$-1.3$ for the UV continuum luminosity functions at $\rm z \le 1$.
The 1500\AA\ continuum luminosity function at $z \sim 3 - 4$
measured from Lyman break galaxies by Steidel et al. (1998) has a
faint end slope of $-1.6$.

The clear trend for the longer wavelength determinations of
the star formation rate to exceed those based on UV continua is
one of the pieces of evidence for significant
extinction at intermediate and high redshifts. The amplitude of
the extinction correction is quite uncertain
Our measurement spans $0.7 < z < 1.9$,
overlapping with the Connolly et al. photometric redshift sample and
allowing a direct comparison between the observed
2800\AA\ luminosity density and that inferred from H$\alpha$.
Our H$\alpha$-based star formation rate is three times larger than the
average of the three redshift bins measured by Connolly et al. (1997).

The star formation rates
derived from line or continuum luminosities depend strongly on the choice of
IMF, evolutionary tracks, and stellar atmospheres that are input into a
specific spectral evolution model. The relevant issue for the present
discussion is the ratio of the star formation rates derived from
H$\alpha$ and the 2800\AA\ continuum. This ratio differs 
significantly for the Scalo and Salpeter IMFs and
is a function of metallicity (Glazebrook et al. 1998). 
Our choice of the Salpeter IMF comes
close to minimizing the difference between the published UV- and our H$\alpha$-derived star
formation rates.
The use of a Scalo IMF and solar metallicity would increase
the apparent discrepancy by a factor of $\sim 2$. The only model considered by
Glazebrook et al. that further reduces the H$\alpha$/2800\AA\ star formation
ratio is the Salpeter IMF with the Gunn \& Stryker (1983) spectral energy
distributions, and this model still leaves us with a factor of $\sim 2$
enhancement in apparent star formation activity measured at H$\alpha$.

If we attribute the entire difference to reddening, the total extinction corrections
at 2800\AA\ and H$\alpha$ are large and model-dependent.
The calculation is sensitive to the relative geometry
of the stars, gas and dust, as well as the adopted reddening curve.
In the extreme case of a homogeneous foreground screen
and a MW or LMC reddening curve,
we derive A$_{2800} = 2.1$~magnitudes.
In local starburst galaxies,
differential extinction between the
nebular gas, and stellar continuum, and scattering produce an effective reddening curve that is
significantly grayer than the MW or LMC curves
(Calzetti, Kinney \&\ Storchi-Bergmann 1994; 1996; Calzetti 1997).
The Calzetti reddening law (Calzetti 1997)
is appropriate for geometries in which the
stars, gas and dust are well mixed.
In this model, our estimate of the dust extinction
at 2800\AA\ is one to two magnitudes larger than in the simple screen case, and
is an uncomfortably large correction compared to results from other methods.

\section{Extremely Red Objects (EROs)}

Several groups have discovered a population of galaxies with 
extremely red colors R$-$K $> 5$ or 6.
However, the statistics of EROs is still very poor and
the nature of these objects remains unclear. The central issue is to 
understand whether EROs are intrinsically red stellar
systems formed at high redshifts in a monolithic collapse or 
highly reddened starburst galaxies at low to moderate redshifts. 
The detection of strong sub-mm continuum from ERO HR~10
(Dey et al. 1999; Cimmati et al. 1998) provides conclusive evidence
that some of EROs, if not all, are indeed dusty, starburst galaxies at
a star formation rate of $\rm 500-1000 M_\odot~yr^{-1}$ at
moderate redshifts ($z \sim 1 -2$).

We have obtained deep ground-based optical images of 27 NIC3 fields, yielding
$\sim$ 20 square arcminutes. Among these fields, we have identified about
a dozen of EROs with R$-$H $>5$ and H brighter than 20.6.  The surface density
of EROs with H $<$ 20.6 and R$-$H $>5$ is roughly 0.6/sq. arcminutes.
We also found some evidence that EROs are highly clustered. Among 27 NIC3 fields, 
we found 2 clusters of EROs. Figure 4 shows a cluster
of EROs in a single NIC3 field (0.75$^{''}$) where we also have K-band magnitudes.

\begin{figure}
\plotfiddle{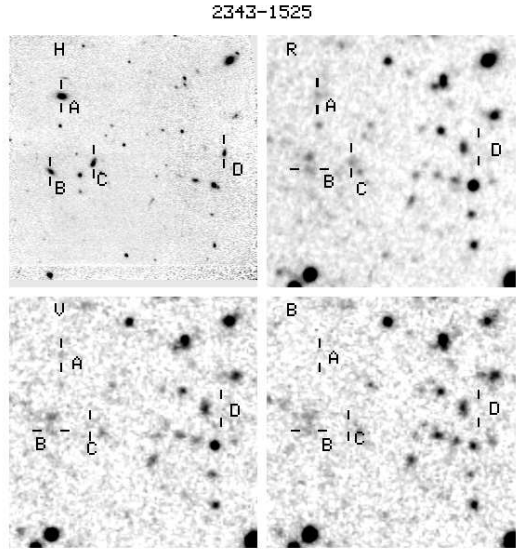}{6.5truecm}{0.0}{50}{50}{-160}{-100}
\caption{This plot shows a cluster of EROs with BVRH images in a single
NIC3 fields (0.75${''}$). All the four EROs have R$-$K $>6$ and
the brightest two A and D have K $\sim 18$. The deep BVR images
were taken with the BTC at CTIO.}
\end{figure}


\acknowledgments

We thank the staff of the Space Telescope Science Institute for their
efforts in making this parallel program possible. This research
was supported, in part, by grants from the Space Telescope Science
Institute, GO-7498.01-96A and P423101.


\begin{references}
\reference Barger, A.J., Cowie, L.L., Sanders, D.B., et al. 1998, Nature, 394, 248
\reference Connolly, A., Szalay, A. S., Dickinson, M., et al. 1997, ApJ, 487, L13
\reference Calzetti, D., Kinney, A.L. \&\ Storchi-Bergmann, T. 1994, ApJ, 429, 582
\reference Calzetti, D. 1997, AJ, 113, 162
\reference Gallego, J., Zamorano, J., Aragon-Salamanca, A., Rego, M. 1995, ApJL, 459, 1
\reference Glazebrook, K., Blake, C., Economou, F., Lilly, S.
et al. 1998, astro-ph/9808276
\reference Hughes, D., et al. 1998, Nature, 394, 241.
\reference Kennicutt, R.C. 1999, ARAA, in press. astro-ph/9807187
\reference Kennicutt, R.C. 1983, ApJ, 272, 54
\reference Lilly, S.J., Le Fevre, O., Hammer, F. \&\ Crampton, D. 1996, ApJ, 460, L1
\reference Lilly, S.J., Eales, S.A., et al. 1999, \apj, in press. astro-ph/9901047
\reference McCarthy, P., Yan, L., 1999, \apj, august issue, in press, astro-ph/9902347
\reference Steidel C., Dickinson, M., et al. 1998, ApJ, in press. astro-ph/9811399
\reference Teplitz, H., et al. 1999, in prep.
\reference Tresse, L., Maddox, S. 1998, \apj, 495, 691
\reference Yan, L. McCarthy, P., 1999, \apjl, July issue, in press, astro-ph/9904427

\end{references}
\end{document}